\documentclass[12pt]{iopart}
\usepackage{amssymb}
\usepackage{txfonts}

\usepackage[dvips]{graphicx}
\usepackage{subfigure}
\usepackage{booktabs}
\usepackage{eurosym}

\usepackage{color}

\begin{document}

\title{Ordering in bidirectional pedestrian flows and its influence on
the fundamental diagram}

\author{J Zhang$^1$, W Klingsch$^1$,  A Schadschneider$^2$ and
A Seyfried$^{3,4}$}

\address{$^1$ Institute for Building Material Technology and Fire
  Safety Science, Bergische Universit\"at Wuppertal,
  Pauluskirchstrasse 11, 42285 Wuppertal, Germany}
\address{$^2$ Institut f\"ur Theoretische Physik, Universit\"at zu
  K\"oln, 50937 K\"oln, Germany}
\address{$^3$ Computer Simulation for Fire Safety and Pedestrian Traffic,
  Bergische Universit\"at Wuppertal, Pauluskirchstrasse 11, 42285
  Wuppertal, Germany}
\address{$^4$ J\"ulich Supercomputing Centre, Forschungszentrum
  J\"ulich GmbH, 52425 J\"ulich, Germany}
\ead{jun.zhang@uni-wuppertal.de, klingsch@uni-wuppertal.de,
as@thp.uni-koeln.de, seyfried@uni-wuppertal.de}

\begin{abstract}
  Experiments under laboratory conditions were carried out to study
  the ordering in bidirectional pedestrian streams and its influence
  on the fundamental diagram (density-speed-flow relation).  The
  Voronoi method is used to resolve the fine structure of the
  resulting velocity-density relations and spatial dependence of the
  measurements. The data show that the specific flow concept is
  applicable also for bidirectional streams. For various forms of
  ordering in bidirectional streams, no large differences among
  density-flow relationships are found in the observed density range.
  The fundamental diagrams of bidirectional streams with different
  forms of ordering are compared with that of unidirectional streams.
  The result shows differences in the shape of the relation for $\rho
  >1.0~$m$^{-2}$. The maximum of the specific flow in unidirectional
  streams is significantly larger than that in all bidirectional
  streams examined.
\end{abstract}

\section{Introduction}

In the last few decades, problems related to bidirectional flow and
its effects on pedestrian dynamics have gained increasing attention.
A large number of models
\cite{Jiang2009,Nagatani2009,Burstedde2001,Hua2008a,Jian2010,Takimoto2002}
have been developed to understand the basic characteristics related
to bidirectional flow including lane formation
\cite{Jian2010,Kretz2006,Hoogendoorn2004}, jamming transition
\cite{Nagatani2009,Takimoto2002,Hua2008,Tajima2002a} and fundamental
diagram \cite{Blue2001}. Computer simulations of various models have
found a jamming transition from a free flow state to a jammed state.
However, this phenomenon has never been observed in experimental as
well as field studies.

\begin{figure*}
\centering\subfigure[Density-velocity]{
\includegraphics[scale=0.75]{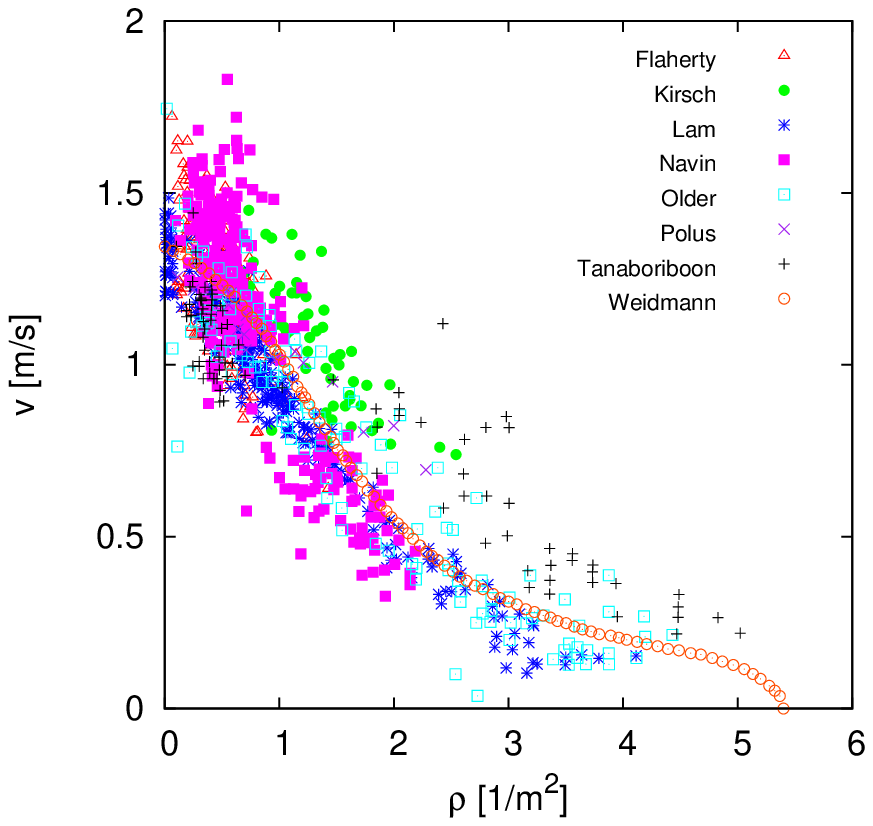}}
\subfigure[Density-specific flow]{
\includegraphics[scale=0.75]{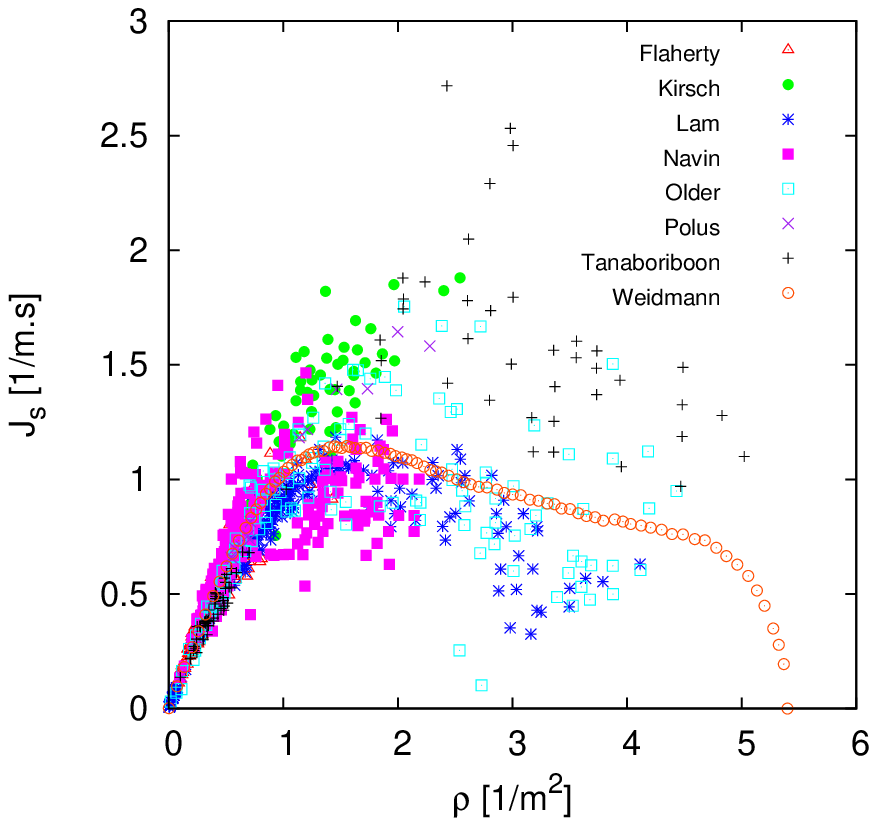}}
\caption{\label{fig-lit} Comparison of fundamental diagrams of
bidirectional pedestrian flow from various studies. Note that the
data given by Weidmann \cite{Weidmann1993}, who performed a
meta-study, is an idealized fundamental diagram obtained by
collecting and fitting 25 other experiments.}
\end{figure*}

Furthermore, a large number of empirical studies have been conducted
to investigate the characteristics of bidirectional flow. Pedestrian
flow characteristics at crosswalks
\cite{Lam2002,Polus1983,O'Flaherty1972,Tanaboriboon1986,Navin1969}
and shopping streets \cite{Older1968} were studied in different
countries. Some of the studies compare mean walking speed and
maximum flow rate among different regions. The influence of facility
width on the pedestrian characteristics is studied in
\cite{Tanaboriboon1986,Older1968}. FIG.~\ref{fig-lit} assembles the
fundamental diagrams of bidirectional flow from these empirical
studies. From the density-velocity relationship in
FIG.~\ref{fig-lit}(a) it can be seen that they follow nearly the
same trend. Further, we compare the relationship between density and
specific flow $J_s $, as shown in FIG.~\ref{fig-lit}(b), using the
hydrodynamic relation $J_s = \rho \cdot v$. In this graph,
differences can be observed especially for densities $\rho
>2.0~$m$^{-2}$. The density values where the specific flow reach the
maximum range from about $1.3~$m$^{-2}$ to $2.3~$m$^{-2}$. Also the
maximum specific flows from different studies range from about
$1.0~($m$\cdot$s$)^{-1}$ to $2.0~($m$\cdot$s$)^{-1}$. But from these
studies no conclusion could be drawn at which density the flow
reaches zero due to congestion.

\begin{figure*}
\centering\subfigure[Density-velocity]{
\includegraphics[scale=0.8]{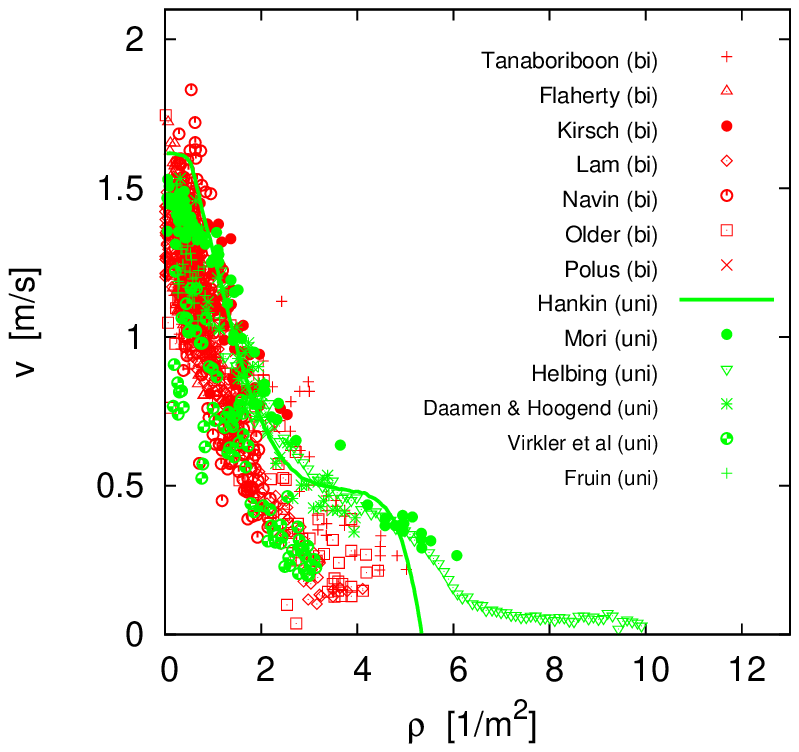}}
\subfigure[Density-specific flow]{
\includegraphics[scale=0.8]{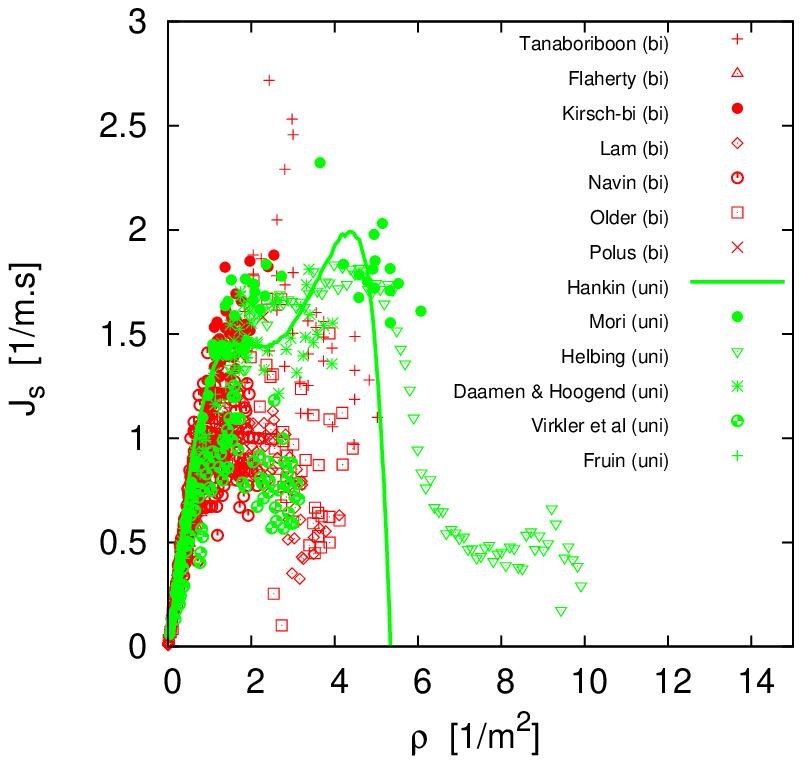}}
\caption{\label{fig-lit-com} The fundamental diagram of uni- and
bidirectional pedestrian flow from different previous studies. }
\end{figure*}

On the other hand, there is still no consensus whether the
fundamental diagrams of uni- and multi-directional flows are
different or not. Predtechenskii and
Milinksii~\cite{Predtechenskii1978} and Weidmann \cite{Weidmann1993}
neglected the differences in accordance with Fruin, who states that
the fundamental diagrams of multi- and uni-directional flow differ
only slightly \cite{Fruin1971}. This disagrees with results of Navin
and Wheeler \cite{Navin1969} who found a reduction of the flow in
dependence of directional imbalances. Further, Pushkarev et
al.~\cite{Pushkarev1975} and Lam et al.~\cite{Lam2002,Lam2003}
assume that bidirectional flow is not substantially different from
unidirectional flow as long as the densities of the opposite streams
are not too different. However, Older et al.~state that different
ratios of flow in bidirectional stream do not show any consistent
effect on the walking speed \cite{Older1968}. Besides, Helbing et
al.~\cite{Helbing2005} concluded that counterflows are significantly
more efficient than unidirectional flows. However, they compare
average flow values without considering the influence of the
density. Kretz et al.~\cite{Kretz2006} have reported similar
findings, but the influence of density and variations in time on the
flow are not considered in this study. Facing these disagreements,
we collect the fundamental diagrams of unidirectional
\cite{Hankin1958,Mori1987,Virkler1994a,Helbing2007} and
bidirectional flow in FIG.~\ref{fig-lit-com}. It seems that the
fundamental diagrams of unidirectional flow lie above those of
bidirectional flow, especially for $\rho > 1.0~$m$^{-2}$. Since the
data compared are obtained under different experimental situations
and different measurement methods, thus we could not conclude
whether and how the type of flow (uni- or bidirectional) influences
the fundamental diagram.

The above discussion shows that up to now there is no consensus
about the origin of the discrepancies between different types of
pedestrian flow. In our study, we carry out a series of laboratory
experiments of bidirectional and unidirectional pedestrian streams.
The formation of lanes in bidirectional streams depend on time as
well as space and could be stable or unstable. Moreover the opposing
flows in bidirectional streams could be balanced or unbalanced. To
categorize these types, we classify the bidirectional streams into
{\it Stable
  Separated Lanes (SSL)} and {\it Dynamical Multi-Lanes (DML) flow}.
According to the typical densities in the opposing streams we
introduce the types {\it Balanced Flow Ratio (BFR)} and {\it
Unbalanced Flow Ratio (UFR)}. The aim of our study is to obtain the
fundamental diagrams and calibrate the previous research results. We
will study the effects of different corridor widths and different
flow types on the fundamental diagrams of bidirectional streams. We
also compare the fundamental diagram of unidirectional and
bidirectional streams.

The structure of the paper is as follows. In Section~\ref{sec-setup}
we describe the setup of experiment. The identification of lanes is
studied by means of the Voronoi diagram in Section~\ref{sec-lane}.
Section~\ref{sec-funda} compares the fundamental diagrams for
different stream types. Finally, the conclusions from our
investigations will be discussed.

\section{Experiment Setup}
\label{sec-setup}

The experiments were performed in hall 2 of the fairground
D\"usseldorf (Germany) in May 2009. Up to 350 participants, mostly
students, participated in the experiments. Each of them was paid 50
{\euro} per day. The mean age and height of the participants was $25
\pm 5.7$ years and $1.76 \pm 0.09$~m respectively. The average free
velocity $v_0= 1.55 \pm 0.18$~m/s was obtained by measuring the free
movement of 42 participants.

\begin{figure*}
\centering{
\includegraphics[scale=0.7]{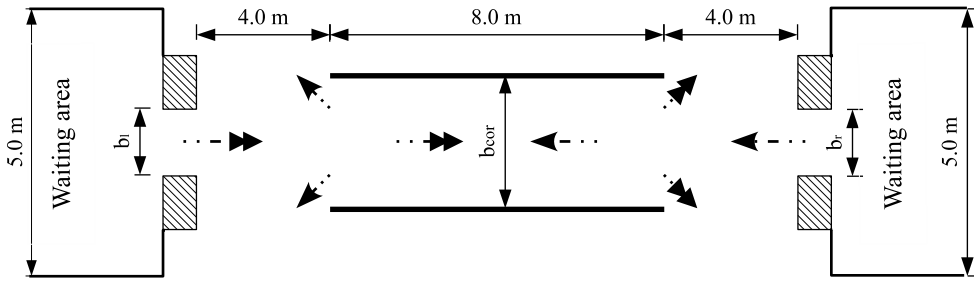}}
\caption{\label{fig1} Schematic illustration of the bidirectional
pedestrian experiment in a corridor. }
\end{figure*}

\begin{figure*}
\centering\subfigure[BFR-SSL]{
\includegraphics[scale=0.9]{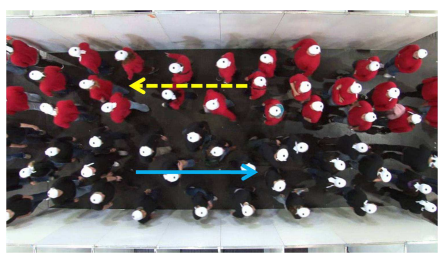}
\includegraphics[scale=0.55]{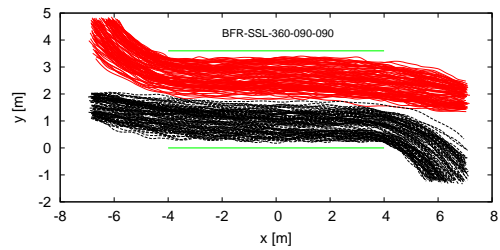}}
\subfigure[BFR-DML]{
\includegraphics[scale=1.0]{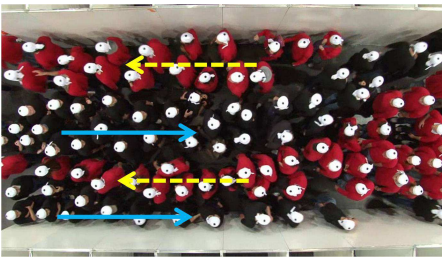}
\includegraphics[scale=0.55]{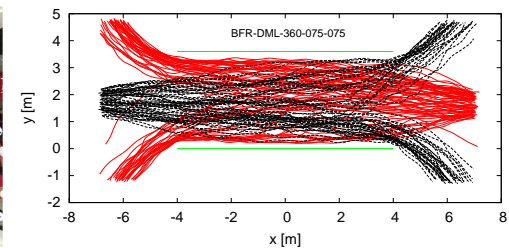}}
\subfigure[UFR-DML]{
\includegraphics[scale=0.9]{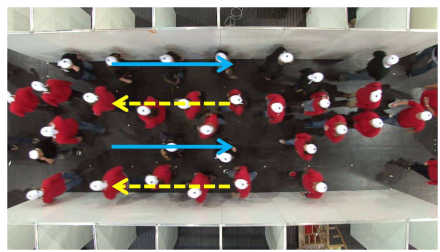}
\includegraphics[scale=0.55]{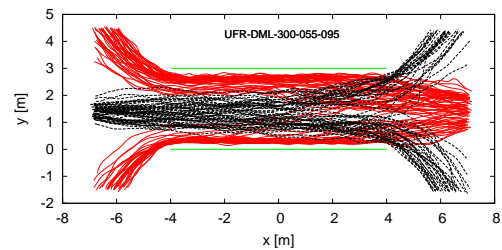}}
\caption{\label{fig2} Snapshots from the experiments (left) and
actual trajectories of pedestrians (right). In the plots, red solid
lines represent the paths of left moving pedestrians, while the
black dashed lines represent the paths of right moving pedestrians.}
\end{figure*}

FIG.~\ref{fig1} shows a sketch of the experimental setup. 22 runs of
bidirectional pedestrian streams (see
TABLE.~\ref{table1}--\ref{table3}) were performed in straight
corridors with widths of 3.0~m and 3.6~m, respectively. To control
the density inside the corridor and the ratio of the opposing
streams, the width $b_{l}$ of the left entrance and $b_{r}$ of the
right entrance were changed in each run (for details, see
FIG.~\ref{fig1} and TABLE.~\ref{table1}--\ref{table3}). Before the
start of an experiment, the participants were arranged within the
waiting area at the left and right side of the corridor. At the
beginning of each run, the pedestrian pass through a 4~m passage
into the corridor. The passage was used as a buffer to minimize the
effect of the entrance.  In this way, the flow in the corridor was
nearly homogeneous over the entire width of the corridor. When a
pedestrian arrived at the other side of the corridor, he or she left
the corridor from the passage and returned to the waiting area for
the next run. To vary the form of the ordering the test persons get
different instructions and the width of entrances is changed. These
variations result in different types of bidirectional flow:

\emph{BFR-SSL flow} (FIG.~\ref{fig2}(a)): This type of flow was
realized in our experiments by using the same entrance width for
both directions ($b_l$ = $b_r$) and giving no instruction to the
test persons about which exit they have to choose. The opposing
flows segregate and occupy proportional shares of the corridor.
Stable lanes formed autonomously and immediately after the run
starts.

\emph{BFR-DML flow} (FIG.~\ref{fig2}(b)): Again identical widths
$b_l$ and $b_r$ are chosen in the experiments, but the instruction
to the test person at the beginning of the experiments changed. The
participants were asked to choose an exit at the end of the corridor
according to a number given to them in advance (odd numbers exit to
the left, even numbers to the right). With this initial condition
again lane formation is observable, but the lanes are unstable and
vary in time and space. This type of flow is comparable with two
stream crossing at a small angle.

\emph{UFR-DML flow} (FIG.~\ref{fig2}(c)): In this case the widths of
entrances $b_l$ and $b_r$ are different and the participants are
instructed to choose an exit at the end of the corridor according to
a number given at the beginning of the day (odd numbers left, even
numbers to the right). Again lanes are unstable and vary in time and
space. The cumulated trajectories indicate that the flow ratio of
the opposing streams is unbalanced.

The experiments were recorded by two cameras mounted on a rack at
the ceiling of the hall. To cover the complete region, the left and
the right part of the corridor were recorded separately. The
trajectories were automatically extracted from video recordings
using the software {\it PeTrack} \cite{Boltes2010}. Finally, the
trajectories from the two cameras were corrected and combined. All
the analysis and the pedestrian characteristics below including
flow, density and velocity are obtained from these trajectories.

\section{Lane formation}
\label{sec-lane}

\begin{figure*}
\centering\subfigure[$t=222$~frame $= 13.875$~sec]{
\includegraphics[scale=0.4]{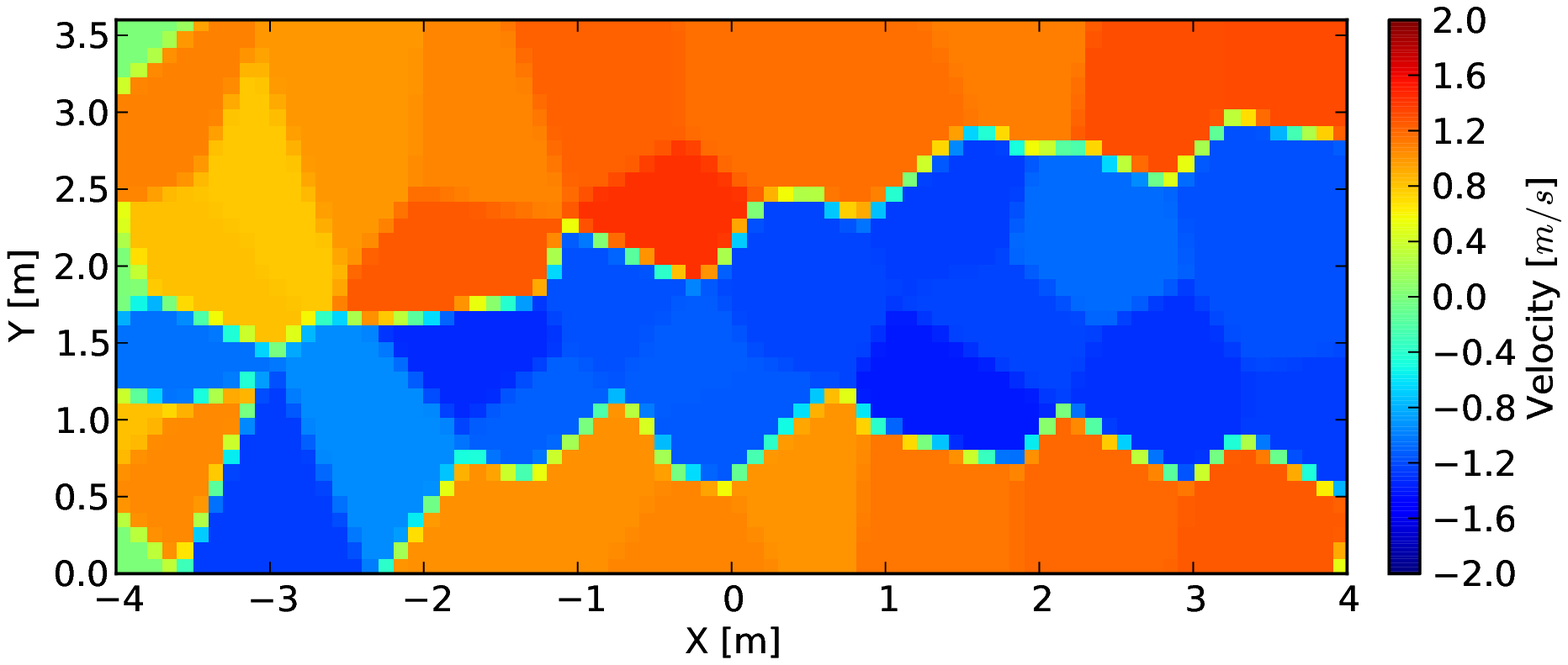}
\includegraphics[scale=0.4]{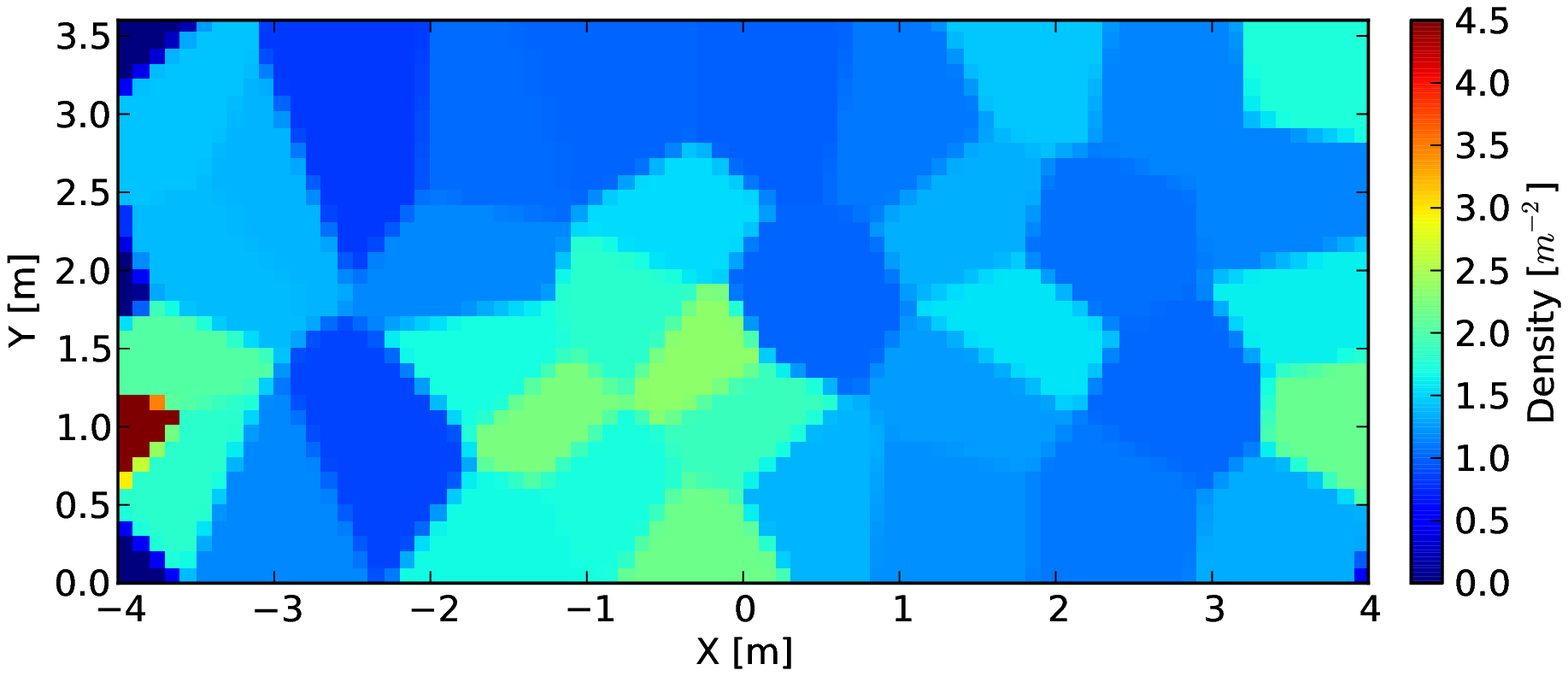}}
\subfigure[$t=822$~frame $= 51.375$~sec]{
\includegraphics[scale=0.4]{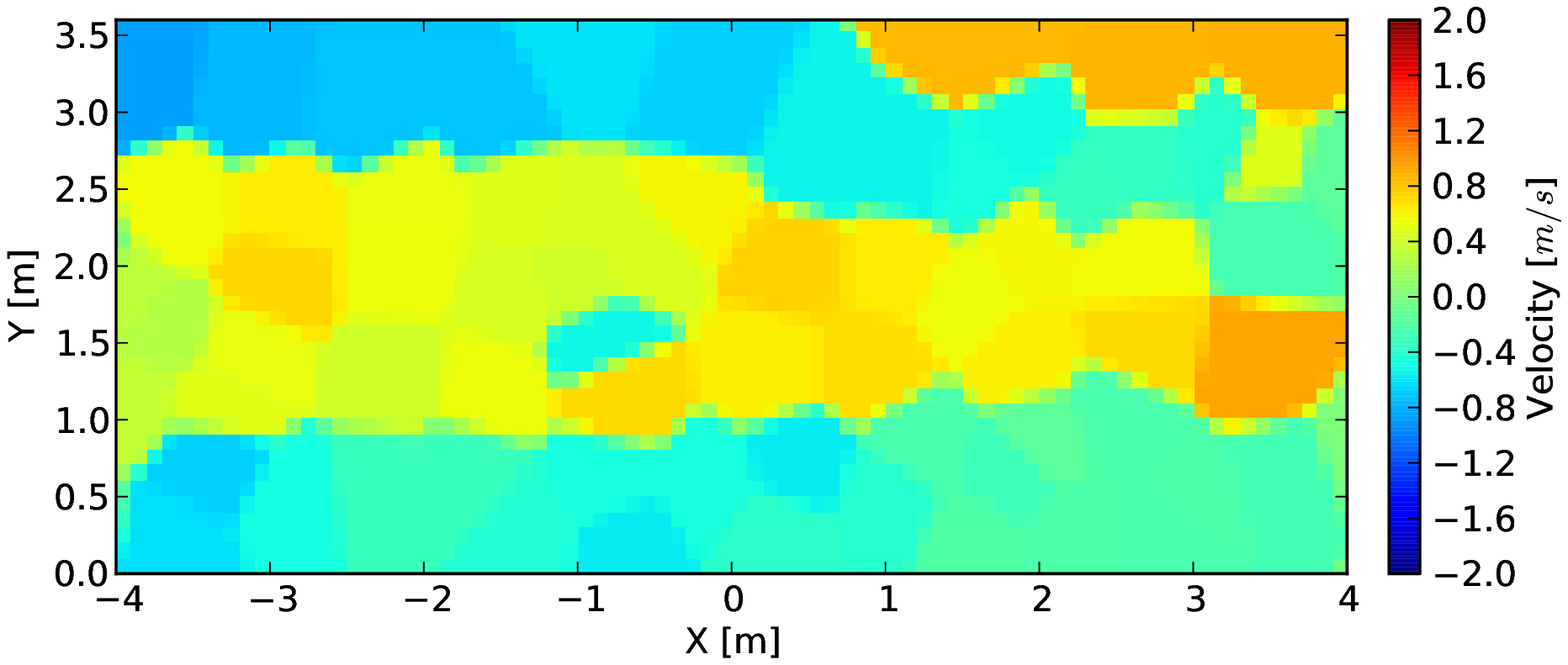}
\includegraphics[scale=0.4]{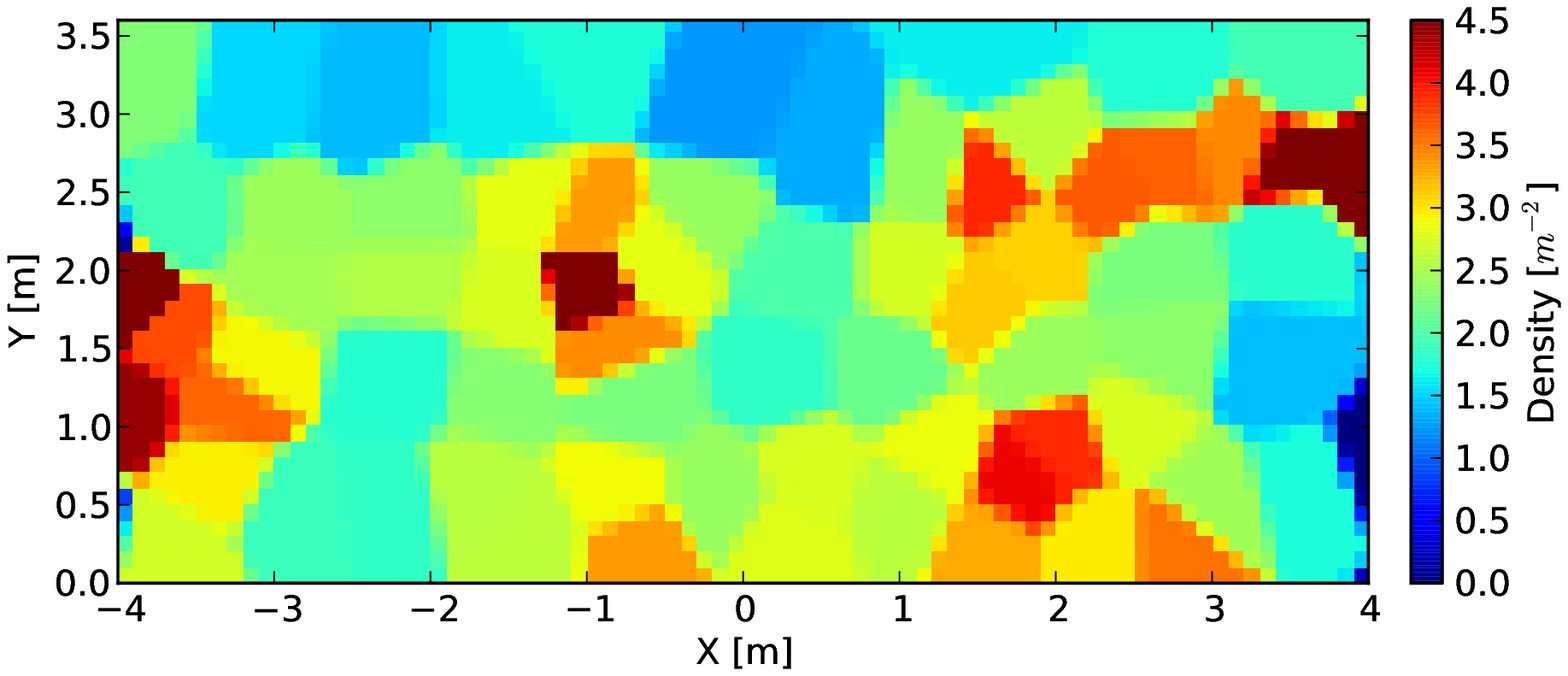}}
\caption{\label{fig7} Velocity and density profiles of experiment
BFR-DML-360-160-160 for different frames (times). }
\end{figure*}

Lane formation, as an important phenomenon in bidirectional flow,
occurs because pedestrians follow closely behind some other person
who moves in the same direction to minimize conflicts with persons
moving in the opposite direction.
Lanes that emerge in this way could be stable (SSL) and unstable
(DML). The recognition and representation of the lanes has been
investigated in different ways, e.g.\ the cluster analysis method
introduced by Hoogendoorn \cite{Hoogendoorn2004}, the bond index
method of Yamori \cite{Yamori1998} et al. as well as the laning
order parameter used to detect lanes in driven colloidal systems
\cite{RexLoewen}. In this study, using the Voronoi method introduced
in \cite{Zhanga,Steffen2010a}, we are able to calculate the
integrated velocity over small ($10~$cm) measurement regions. In
this way, the velocity distribution over the whole space can be
easily obtained. FIG.~\ref{fig7} shows the velocity and density
profiles of BFR-DML-360-160-160 for two different times. We use
different colors to present the value of the velocity and density.
Then the number of lanes in the corridor can be easily determined
from the velocity profiles. The velocity profiles seem to be a good
way to display the lane formation in bidirectional streams. In
comparison, density profiles don't have such ability to show the
lanes clearly. However, it is possible for density profiles to show
some other information such as crowded and dangerous spots.

\section{Fundamental diagram}
\label{sec-funda}

In this study we use the Voronoi method \cite{Zhanga,Steffen2010a}
to analyze the flows quantitatively. We have verified that
alternative methods (see the discussion in \cite{Zhanga}) do not
lead to large changes in the fundamental diagrams. The advantage of
the Voronoi method lies in its small scatter and high resolution in
time and space.

For the analysis a rectangle with a length of $2~$m from $x=-1~$m to
$x=1~$m and the width of the corridor is chosen as the measurement
area. We calculate the densities and velocities every frame
(corresponding to $0.0625~$s) with a frame rate of 16~fps. To
determine the fundamental diagram only data from stationary flows
are used which are selected manually by analyzing the time series of
density and velocity. Finally, we use one frame per second to limit
the number of data points. The influences of corridor width and the
type of flow on the fundamental diagram are studied and compared
with the diagram of unidirectional flow.

\subsection{Influence of corridor width}\label{subsec-width}

\begin{figure*}
\centering\subfigure[Density-velocity]{
\includegraphics[scale=0.8]{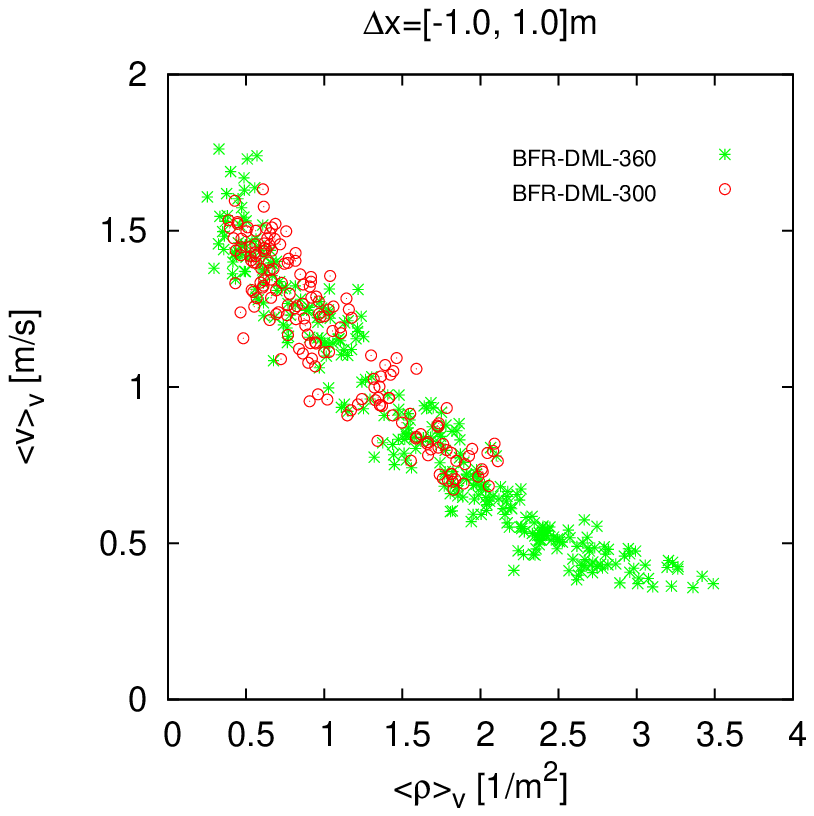}}
\subfigure[Density-specific flow]{
\includegraphics[scale=0.8]{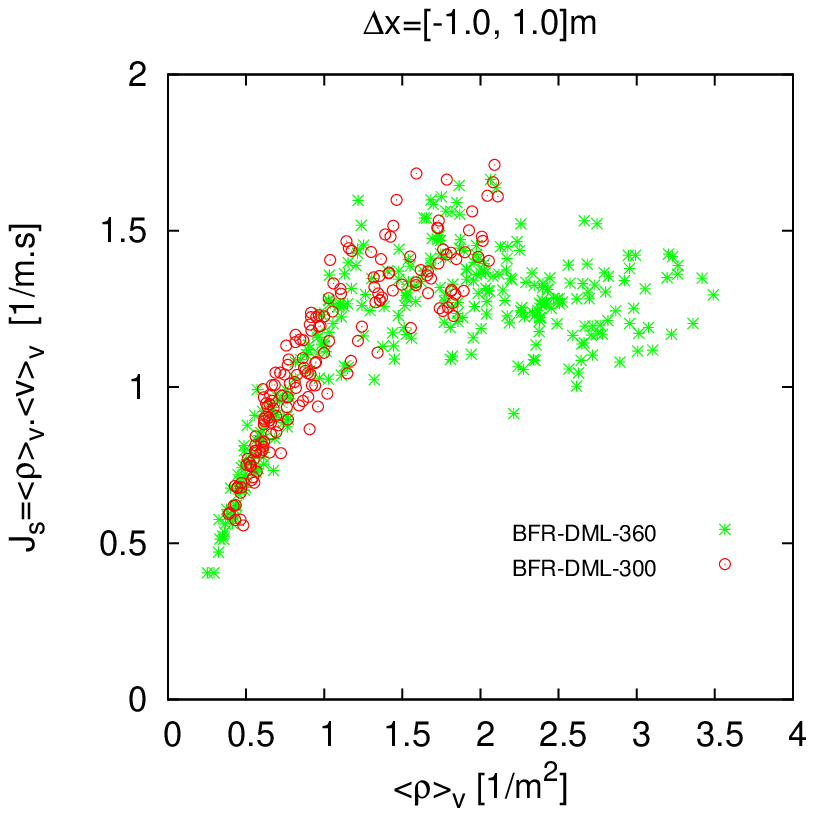}}
\caption{\label{fig3} Comparison of fundamental diagrams of DML flow
at different corridor widths. }
\end{figure*}

Firstly, we study the influence of the corridor width on the
fundamental diagram. For BFR-DML flow, two widths $b_{\rm
  cor} = 3.0~$m and $b_{\rm cor} = 3.6~$m are chosen in the
experiments and the fundamental diagrams are compared. Due to the
limitation of the time resources for the experiment only densities
$\rho < 2.0~$m$^{-2}$ are realized for $b_{\rm cor} = 3.0~$m. As
shown in FIG.~\ref{fig3}, the fundamental diagrams are in good
agreement for different widths. FIG.~\ref{fig3}(b) shows the
relationship between the Voronoi density and specific flow for these
two widths. The specific flow reaches its maximum of
$1.5~$(ms)$^{-1}$ at a density of $\rho = 2.0~$m$^{-2}$.

\subsection{Comparison of SSL and DML flow}

\begin{figure*}
\centering\subfigure[Density-velocity]{
\includegraphics[scale=0.8]{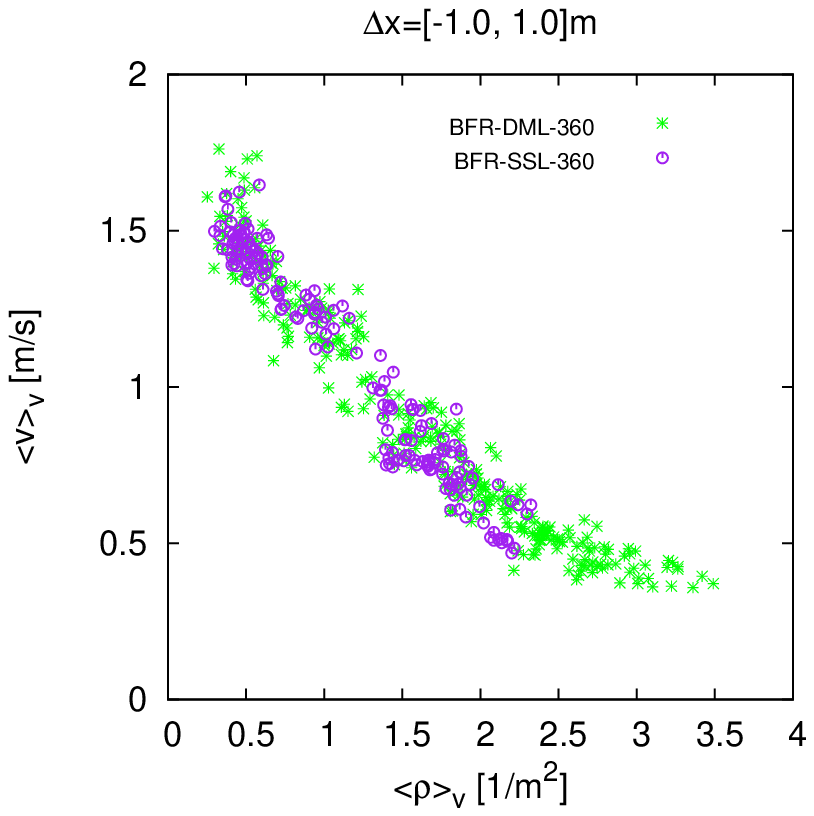}}
\subfigure[Density-specific flow]{
\includegraphics[scale=0.8]{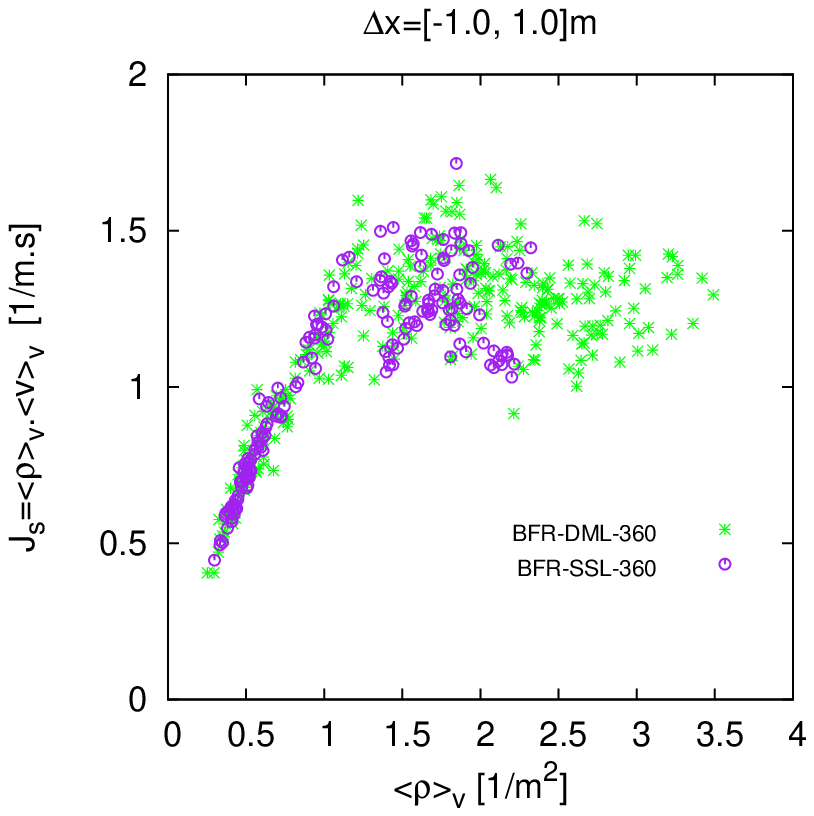}}
\caption{\label{fig4} Comparison of the fundamental diagrams of SSL
and DML flow. }
\end{figure*}

In bidirectional pedestrian flow, especially for DML type, head-on
conflicts and cross-directional conflicts occur.  To investigate
their influence on the flow we compare in FIG.~\ref{fig4} the
fundamental diagram of SSL and DML flow for $b_{\rm cor} = 3.6~$m.
It can be seen that the fundamental diagrams of these two types of
bidirectional flow are consistent at least for the density $\rho <
2.0~$m$^{-2}$. The lower degree of ordering in dynamical multi-lanes
(DML) has no effect on the fundamental diagram which agrees with the
findings of Older \cite{Older1968}.  This might be taken as an
indication that head-on conflicts in multi-lanes have the same
influence as the conflicts at the borders in stable separated lane
flow on the fundamental diagram. On the other hand, the
self-organized lanes increase the order and make pedestrian movement
smoother.  Whether the degree of ordering has an influence on the
fundamental diagram at higher densities can not be decided from our
data. In particular it would be interesting to find out whether the
density where the velocity and specific flow becomes zero depends on
the degree of order.

\subsection{Comparison of BFR and UFR flow}

\begin{figure*}
\centering\subfigure[Density-velocity]{
\includegraphics[scale=0.8]{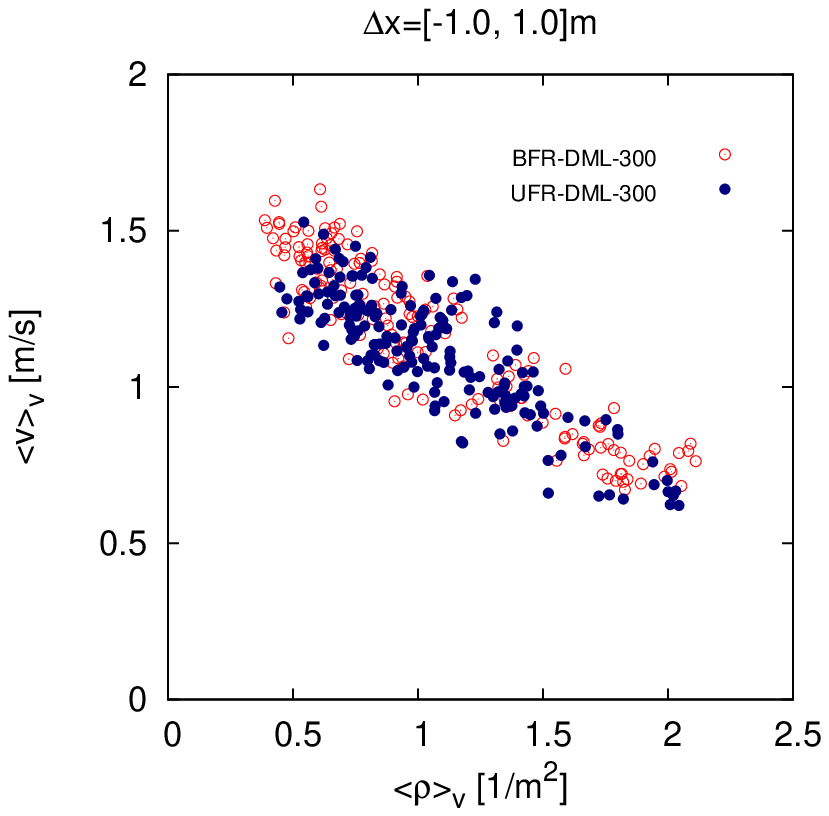}}
\subfigure[Density-specific flow]{
\includegraphics[scale=0.8]{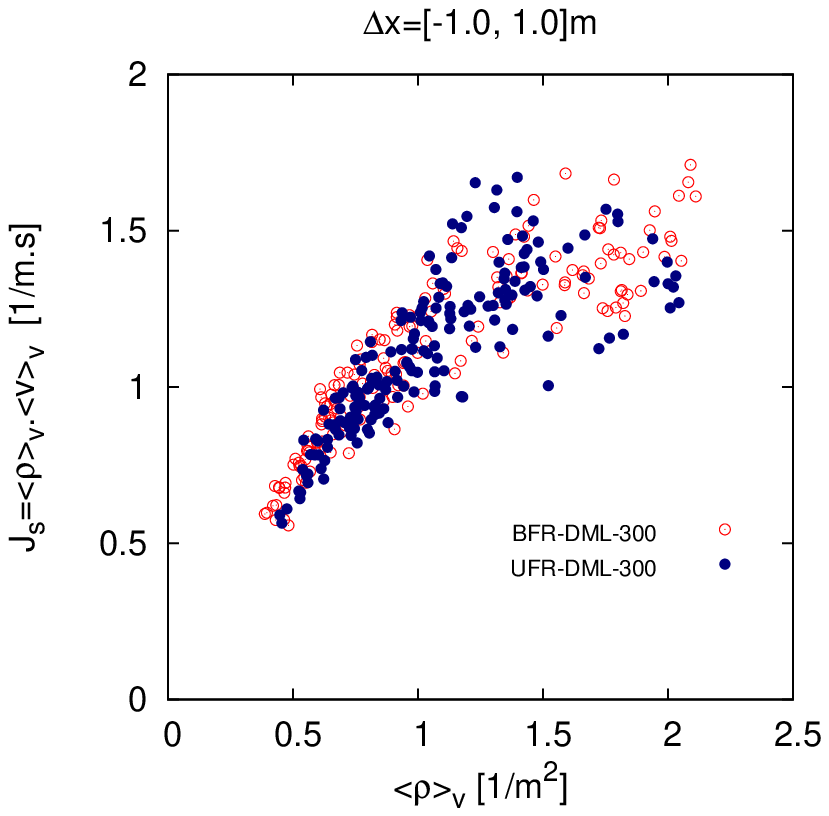}}
\caption{\label{fig5} Comparison of the fundamental diagrams of BFR
and UFR flow. }
\end{figure*}

The flow ratio of the opposing pedestrian streams is another
important factor that is worth studying. Under unbalanced
conditions, pedestrians from the direction with high flow ratio may
dominate and restrain the movement of pedestrians from the opposing
direction. We compare the fundamental diagrams of BFR and UFR flow
to study the influence of the ratio of opposing flows, as shown in
FIG.~\ref{fig5}. This comparison is performed for $b_{\rm cor} =
3.0~$m of DML flow.
Due to the limitation of the number of runs, only data for $\rho <
2.0~$m$^{-2}$ have been obtained.  It can be seen that the asymmetry
in the flows does not affect the fundamental diagrams, at least for
DML flows and densities $\rho < 2.0~$m$^{-2}$.

\subsection{Comparison between uni- and bidirectional flows}

\begin{figure*}
\centering\subfigure[Density-velocity]{
\includegraphics[scale=0.8]{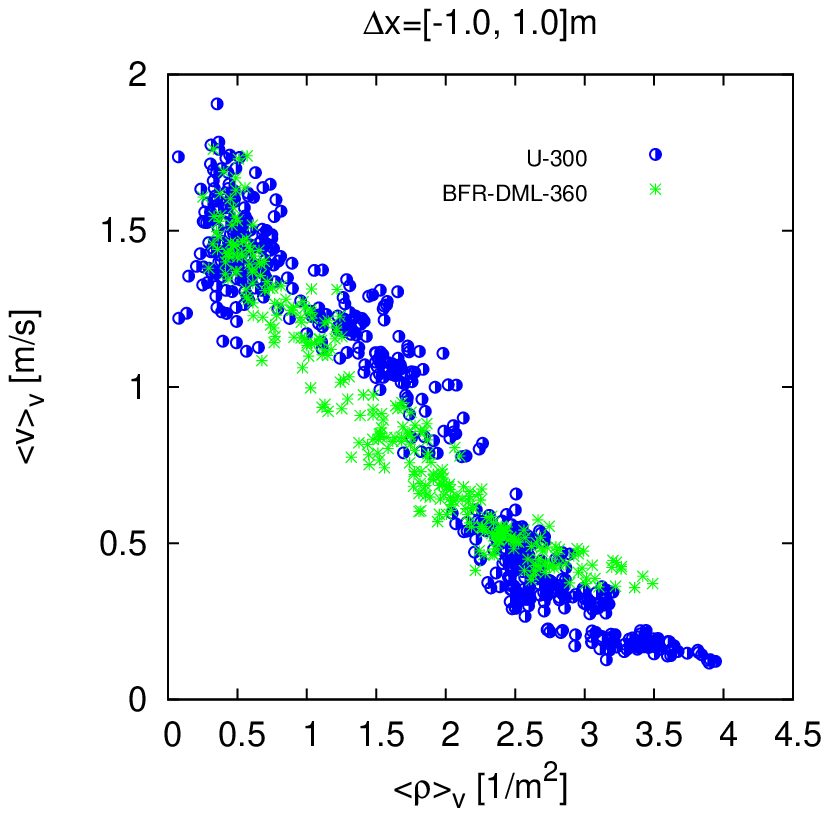}}
\subfigure[Density-specific flow]{
\includegraphics[scale=0.8]{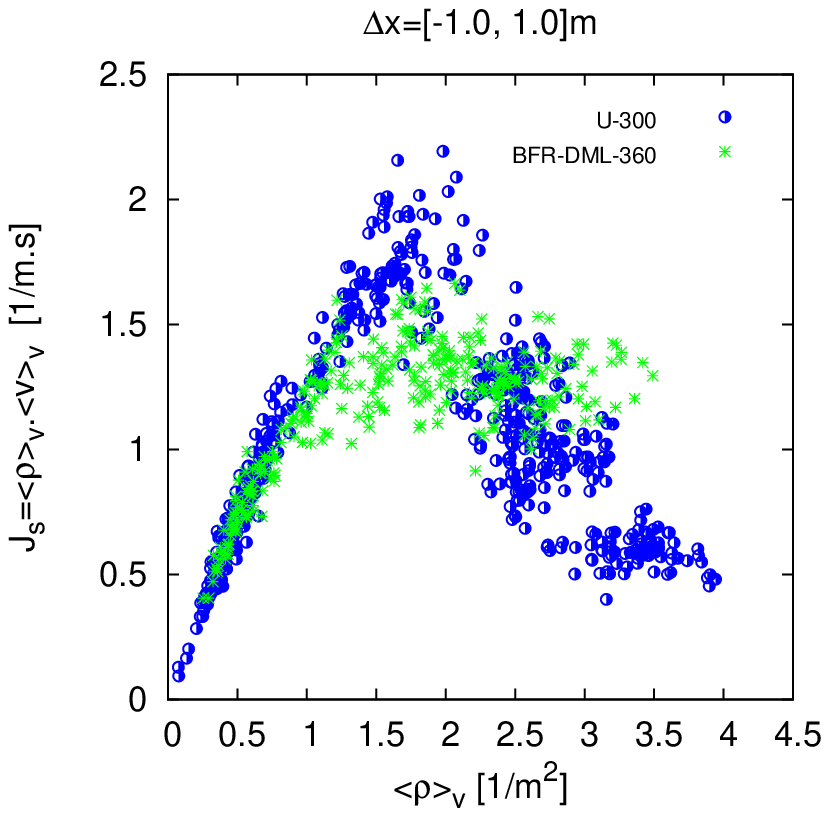}}
\caption{\label{fig6} Comparison of the fundamental diagrams of
unidirectional flow and bidirectional flow. }
\end{figure*}

In this section we compare the fundamental diagram of uni- and
bidirectional flows, see FIG.~\ref{fig6}. Since both the width of
corridor and the forms of bidirectional flow have small influence on
the fundamental diagram, the data for unidirectional flow (U-300) in
\cite{Zhanga} at $b_{\rm cor} = 3.0~$m are used to compare with the
fundamental diagram of BFR-DML flow at $b_{\rm cor} = 3.6~$m. These
experiments were carried out at the same day with the same
participants. It is worth noting that the data of the unidirectional
flow for $\rho > 2.0~$m$^{-2}$ are obtained by slide change of the
experiment setup. To reach densities $\rho > 2.0~$m$^{-2}$ for
unidirectional experiment, a bottleneck at the end of the corridor
is introduced. We discussed in \cite{Zhanga} that the decrease of
the flow is induced by changing the boundary conditions. This limits
the comparability of fundamental diagrams for $\rho > 2.0~$m$^{-2}$.

FIG.~\ref{fig6}(a) shows the relationship between density and
velocity for these two kinds of flow. For the free flow state at
densities $\rho < 1.0~$m$^{-2}$ no significant difference exists.
For $\rho > 1.0~$m$^{-2}$, however, the velocities for
unidirectional flow are larger than that of bidirectional streams.
The difference between the two cases becomes more pronounced in the
flow-density diagram where a qualitative difference can be observed.
In the bidirectional case a plateau is formed starting at a density
$\rho \approx 1.0~$m$^{-2}$ where the flow becomes almost
independent of the density. Such plateaus are typical for systems
which contain 'defects' which limit the flow \cite{JanowskyLebowitz}
and have been observed e.g.\ on bidirectional ant trails \cite{John}
where they are a consequence of the interaction of the ants. In our
experiments the defects are persons moving in the opposite direction
which leads to conflicts and a reduction of the velocity.


The observed difference in the fundamental diagrams implies that SSL
flow should not be interpreted as two unidirectional flows. Although
the self-organized lanes can decrease the head-on conflicts,
interactions between the opposing streams are still relevant.

\section{Conclusion}
\label{sec-conclusion}


Series of well-controlled laboratory pedestrian experiments of
bidirectional flow were performed in straight corridors. Up to 350
persons participated and the whole processes of the experiment were
recorded using two video cameras. The trajectories of each
pedestrian are extracted with high accuracy from the video
recordings automatically using {\em PeTrack}. After comparing the
fundamental diagrams from the four measurement methods introduced in
\cite{Zhanga}, we don't find large difference among these results
and adopt the Voronoi method in this study for its small fluctuation
and high resolution in time and space. For different degree of
ordering in bidirectional pedestrian streams (DML, SSL, BFR and
UFR), no significant differences are found in the fundamental
diagrams for densities below $2.0~$m$^{-2}$. Due to the limitation
of the number of runs, the fundamental diagrams for different types
of bidirectional flow are mainly compared for densities $\rho <
2.0~$m$^{-2}$. Whether a jamming transition, which is due to
overcrowding and lead to a total breakdown of flow, occurs in case
the degree of ordering is lower. This needs to be studied by further
experiments. The fundamental diagrams of bidirectional flow and
unidirectional flow shows clear differences. The maximum flow value
is about $2.0~$(ms)$^{-1}$ for unidirectional flow while
$1.5~$(ms)$^{-1}$ for bidirectional flow. The self-organized lanes
can help to relief the head-on conflicts effectively and increase
the ordering of the stream. However, these conflicts do have
influence on the fundamental diagram of bidirectional flow. It is
also demonstrated that the lanes could be determined easily by
measuring the velocity profile of bidirectional stream using the
Voronoi method.

\begin{table}[htbp]
 \centering\caption{\label{table1}The related parameters in BFR-SSL
 experiments} \lineup
 \begin{tabular}{ccccccc}
 \toprule
  Index & Name & b$_{cor}$ [$m$] & b$_l$ [$m$] & b$_{r}$ [$m$]  & N$_l$ & N$_r$ \\
  \midrule
  1 & BFR-SSL-360-050-050 & 3.60 & 0.50 & 0.50 & 57 & 61 \\
  2 & BFR-SSL-360-075-075 & 3.60 & 0.75 & 0.75 & 56 & 80 \\
  3 & BFR-SSL-360-090-090 & 3.60 & 0.90 & 0.90 & 109 & 105 \\
  4 & BFR-SSL-360-120-120 & 3.60 & 1.20 & 1.20 & 143 & 164 \\
  5 & BFR-SSL-360-160-160 & 3.60 & 1.60 & 1.60 & 143 & 166 \\
  \bottomrule
 \end{tabular}
\end{table}

 \begin{table}[htbp]
 \centering\caption{\label{table2}\centering The related parameters in BFR-DML
 experiments} \lineup
 \begin{tabular}{ccccccc}
 \toprule
  Index & Name & b$_{cor}$ [$m$] & b$_l$ [$m$] & b$_{r}$ [$m$]  & N$_l$ & N$_r$ \\
 \midrule
  1 & BFR-DML-300-050-050 & 3.00 & 0.50 & 0.50 & 54 & 71 \\
  2 & BFR-DML-300-065-065 & 3.00 & 0.65 & 0.65 & 64 & 83 \\
  3 & BFR-DML-300-075-075 & 3.00 & 0.75 & 0.75 & 61 & 86 \\
  4 & BFR-DML-300-085-085 & 3.00 & 0.85 & 0.85 & 119 & 97 \\
  5 & BFR-DML-300-100-100 & 3.00 & 1.00 & 1.00 & 125 & 105 \\
  6 & BFR-DML-360-050-050 & 3.60 & 0.50 & 0.50 & 56 & 74 \\
  7 & BFR-DML-360-075-075 & 3.60 & 0.75 & 0.75 & 62 & 65 \\
  8 & BFR-DML-360-090-090 & 3.60 & 0.90 & 0.90 & 110 & 102 \\
  9 & BFR-DML-360-120-120 & 3.60 & 1.20 & 1.20 & 115 & 106 \\
  10 & BFR-DML-360-160-160 & 3.60 & 1.60 & 1.60 & 140 & 166 \\
  11 & BFR-DML-360-200-200 & 3.60 & 2.00 & 2.00 & 143 & 166 \\
  12 & BFR-DML-360-250-250 & 3.60 & 2.50 & 2.50 & 141 & 163 \\
  \bottomrule
 \end{tabular}
\end{table}

 \begin{table}[htbp]
 \centering\caption{\label{table3}\centering The related parameters in UFR-DML
 experiments}\lineup
 \begin{tabular}{ccccccc}
  \toprule
  Index & Name & b$_{cor}$ [$m$] & b$_l$ [$m$] & b$_{r}$ [$m$]  & N$_l$ & N$_r$ \\
  \midrule
  1 & UFR-DML-300-050-070 & 3.00 & 0.50 & 0.70 & 72 & 63 \\
  2 & UFR-DML-300-050-085 & 3.00 & 0.50 & 0.85 & 61 & 64 \\
  3 & UFR-DML-300-055-095 & 3.00 & 0.55 & 0.95 & 58 & 70 \\
  4 & UFR-DML-300-065-105 & 3.00 & 0.65 & 1.05 & 117 & 112 \\
  5 & UFR-DML-300-080-120 & 3.00 & 0.80 & 1.20 & 116 & 103 \\
  \bottomrule
 \end{tabular}
\end{table}


\end{document}